\documentstyle[12pt]{article}

\makeatletter
\@addtoreset{equation}{section}
\makeatother

\begin{document}

\vspace*{-10mm}

\baselineskip12pt
\begin{flushright}
\begin{tabular}{l}
{\bf KEK-TH-444 }\\
{\bf KEK Preprint 95-85}\\
July 1995
\end{tabular}
\end{flushright}

\baselineskip18pt
\vspace{8mm}
\begin{center}
{\Large \bf Time reversal violation in $K^+ \rightarrow
\mu^+ \nu \gamma$
decay and three Higgs model} \\

\vglue 8mm
{Makoto Kobayashi$^*$, Ting-Ting Lin$^{*,\dagger}$
\footnote{Present address, Physics Department, National University
of Singapore, 10 Kent Ridge Crescent, Singapore 0511.}
 and Yasuhiro Okada$^*$}\\
\vglue 4mm
{\it Theory Group, KEK, Tsukuba, Ibaraki 305, Japan$^*$ }\\
and\\
{\it Physics Department, Xiamen University, Xiamen, P.R.
China$^{\dagger}$ }\\
\vglue 20mm
{\bf ABSTRACT} \\
\vglue 10mm
\begin{minipage}{14cm}
\baselineskip12pt
Transverse muon polarization in the $K^+ \rightarrow \mu^+ \nu \gamma$
decay is calculated in the model with scalar and pseudo scalar
four-Fermi interactions.  Combined with a similar calculation in the
$K^+ \rightarrow \mu^+ \nu \pi^0$ decay, a possible constraint
on parameters in the three-Higgs model is obtained assuming
sensitivity of the up-coming KEK experiment. It is pointed out
that the predictions for the two polarizations are strongly
correlated in the three Higgs model.
\end{minipage}
\end{center}
\vfill
\newpage
\section{Introduction}
{}~~\\

Although the present experimental knowledge on CP violation is
consistently explained by a simple phase of the quark flavor
mixing matrix, it may not be the only source of the CP violation.
In fact, some types of physics beyond the standard model contain new
physical phases which could induce different kinds of CP and T
violating interactions.  Therefore, it is important to search for
such interactions in various processes.

Measurements of transverse muon
polarization in the $K^+ \rightarrow \mu^+\nu \pi ^{0} $
process have received attentions as a process to probe a T
violating interaction\cite{Sakurai}.
Although the standard model prediction to the polarization is too
small to be measured in the near future, possible extensions like
multi Higgs models can induce measurable effects\cite{Kpimunu}.

A new experiment is under preparation at KEK aiming at  measuring
the transverse polarization of muon ($P_T$) up
to the level of 5 $\times 10^{-4}$ \cite{Kuno}.
This would be an improvement by a factor 10 from the present
experimental bound which is $(-3.1\pm5.3)\times 10^{-3}$ \cite{Blatt}.

In the same experiment, the transverse muon polarization of the
$K^+ \rightarrow \mu^+ \nu \gamma$ decay will be also measured.
Since this is proportional to
($\vec{p}_\mu \times \vec{p}_\gamma) \cdot \vec{s}_\mu$
where $\vec{p}_\mu$ and $\vec{p}_\gamma$ are momenta
of muon and photon and $\vec{s}_\mu$ is spin of muon, this quantity
changes its sign under time reversal operation.
Therefore, the measurement of the transverse muon polarization
in this process will also give us useful information
on possible new sources of T and CP violating interactions
\cite{Marciano}. For example, in Ref.\cite{tensor} this polarization
was considered as a probe to possible tensor interactions
in the kaon decay.
Although the transverse muon polarization from the CP violation
in the standard model is negligible, the electromagnetic final state
interaction can mimic the T violation effects which are estimated
to be as large as $10^{-3}$\cite{Geng}. This is in contract with
the $K^+ \rightarrow \mu^+\nu \pi ^{0} $
process where the final
state interaction can only produce the effect of  $10^{-6}$
\cite{FSI}. Since the sensitivity for the polarization
measurement in this mode is expected to be similar to that
of the $K^+ \rightarrow \pi^0\mu^+\nu$ mode at the coming
experiment\cite{Kuno},
we will be able to search for the T violating effects below
$10^{-3}$ level if we can properly subtract the contributions
from the final state interaction.

In this paper, we consider prediction of the transverse muon
polarization in the $K^+ \rightarrow \mu^+\nu\gamma$ process
as well as the $K^+ \rightarrow \pi^0\mu^+\nu$ process in multi
Higgs models.
In these models, new scalar and pseudo scalar four Fermi interactions
are induced from exchange of charged Higgs bosons and these
interactions contain new physical phases.
We will determine how the measurements of transverse muon polarization
for these two processes put constraints on these new interactions.
In particular, we will consider a three Higgs model and show that
the predictions for the above two processes are strongly correlated
after taking account of other phenomenological constraints.
Therefore, it is very important to measure the transverse muon
polarization in both processes to clarify the nature of possible
CP violating effects in the three Higgs model.

This paper is organized as follows.  In section 2, we will introduce
scalar and pseudo scalar interaction and calculate the decay rate
and transverse muon polarization for $K^+ \rightarrow \mu^+\nu\gamma$
and $K^+ \rightarrow \pi^0\mu^+\nu$ processes.
In section 3, we consider a three Higgs doublet model and obtain
constraints on Higgs coupling constants
by these two processes. Discussions on the results are given
in section 4. In the appendix A,
the  $K^+ \rightarrow \mu^+\nu\gamma$
amplitude due to the pseudo scalar interaction is derived.
The appendix B contains several functions for the branching
ratio and polarization calculations.

\section{Muon Polarization in
$K^+ \rightarrow \mu^+\nu\gamma$
and $K^+ \rightarrow \mu^+\nu\pi^0$ Decay}

In this section we will present calculations of transverse muon
polarization
and decay rates for $K^+ \rightarrow \mu^+\nu\gamma$ process.
For completeness, we also give results of a similar calculation for
$K^+ \rightarrow \pi^0\mu^+\nu$ process.

We start from the following four-fermi interaction,

\begin{eqnarray}
{\cal L}  & = & - \frac{G_{F}}{\sqrt{2}} \sin \theta
_{c} \bar{s} \gamma_\mu (1 - \gamma_5) u \bar{\nu} \gamma^\mu
(1-\gamma_5) \mu \nonumber\\
&  & + G_S \bar{s}u \bar{\nu} \frac{1+ \gamma_5}{2} \mu
+ G_P \bar{s} \gamma_5 u \bar{\nu} \frac{1+ \gamma_5}{2} \mu
\nonumber\\
&  & + h.c.,
\label {eq:lag}
\end{eqnarray}
where $G_F$ is  the Fermi constant and $\sin \theta_{c}$ = 0.22.
We have introduced two coupling constants $G_S$ and $G_P$.
These constants are in general complex.
In this section, we treat $G_S$ and $G_P$ as new coupling constants.
Later, when we consider the multi Higgs models, these terms are
supposed to be induced from the charged Higgs exchange and
$G_S$ and $G_P$ are expressed as functions of charged Higgs
masses and coupling constants in the multi Higgs model.

The $K^+ \rightarrow \mu^+\nu\gamma$ amplitude in the standard
model can be divided into two parts. i.e., internal bremsstrahlung
($M_{IB}$)
and structure dependent ($M_{SD}$) terms\footnote{ Detailed account
for the radiative semileptonic kaon decay within the standard model is
found in Ref. \cite{Bijnens}}.

\begin{eqnarray}
M_{V-A}  & = & M_{IB} + M_{SD},\\
M_{IB}   & = & -ie  \frac{G_F}{\sqrt{2}} \sin \theta_{c} \sqrt{2}
f_K m_\mu \epsilon_\nu^*(q)K^\nu,\\
M_{SD}     & = & ie  \frac{G_F}{\sqrt{2}} \sin \theta_{c} L_{\nu}
\epsilon_\mu^*(q) H^{\mu\nu},
\end{eqnarray}
where

\begin{eqnarray}
L^\nu & = & \bar{u}(k) \gamma^\nu (1-\gamma_5)\upsilon(\ell),\\
H^{\mu\nu} & = & \frac{A}{m_K}p \cdot q(-g^{\mu\nu} +
\frac{p^\mu q^\nu}
{p \cdot q}) + i \frac{V}{m_K}\epsilon^{\mu\nu\alpha \beta}
q_\alpha p_\beta,\\
K^\mu & = & \bar{u}(k)(1+\gamma_5)( \frac{p^\mu}{p \cdot q}-
\frac{q\cdot \gamma \gamma^\mu + 2 \ell^\mu}{2 \ell \cdot q})
\upsilon (\ell).
\end{eqnarray}
Here $p^\mu$, $q^\mu$, $\ell^\nu$, $k^\nu$ are the $K^+$, photon,
muon and neutrino four momenta, respectively
and $\bar{u}(k)$ and $\upsilon(\ell)$
are neutrino and muon wave functions. $ \epsilon_\nu$ is the
photon polarization vector. The kaon decay constant
$f_K$ is defined as,

\begin{eqnarray}
< 0  | \bar{s} \gamma^\mu \gamma^5 u | K^+(p)> = -i\sqrt{2}f_K
p^\mu,
\end{eqnarray}
and V and A are defined as follows,

\begin{eqnarray}
& & \int d^4 x e^{iqx}<0 | T
(\bar{s} \gamma^\nu \gamma^5 u(0) J_{em}^{\mu}(x))| K^+(p)>\nonumber \\
& = & -\sqrt{2} f_K( g^{\mu\nu} + \frac{p^{\mu} (p-q)^\nu}{p\cdot q})
+ \frac{A}{m_k}p\cdot q(g^{\mu\nu} - \frac{p^{\mu} q^\nu}{p\cdot q}),\\
& & \int d^4 x e^{iqx}<0 | T
(\bar{s} \gamma^\nu u(0) J_{em}^{\mu}(x))| K^+(p)> \nonumber\\
& = & i \frac{V}{m_K} \epsilon^{\mu\nu\alpha\beta}q_\alpha p_\beta,
\end{eqnarray}
where $J_{em}^{\mu}(x)$ is the electromagnetic current. Note that
the form factors V and A are real since CP is conserved in the
strong interaction.

Let us consider the effects of the scalar and pseudo scalar couplings.
In the appendix A, we show that only the pseudo scalar coupling
can contribute to this process and that the amplitude induced by
the $G_P$
coupling constant is proportional to $M_{IB}$, so that no new form
factor
is necessary. This is quite different from the case of the tensor
interaction where a new form factor should be introduced\cite{tensor}.
The amplitude is given by,

\begin{equation}
M_P  =  -ie\frac{G_P}{2}\frac{\sqrt{2}f_{K}m_{K}^{2}}{m_s + m_u}
\epsilon_\mu^*(q)K^\mu,
\label{eq:M_P}\\
\end{equation}
and $m_s$ and $m_u$ are the strange and up quark masses. Combining
two expression, the total amplitude becomes

\begin{eqnarray}
M & = & M_{IB} + M_{SD} + M_{P}\nonumber \\
& = & -ie \frac{G_F}{\sqrt{2}}(1 + \Delta_P) \sin \theta_{c}
\sqrt{2}f_K m_\mu \epsilon^{*}_{\mu} (q) K^\mu \nonumber\\
& & + ie \frac{G_F}{\sqrt{2}} \sin \theta_c L_\nu
\epsilon^{*}_{\mu} (q) H^{\mu\nu},
\end{eqnarray}
where

\begin{equation}
\Delta_P = \frac{G_P}{\sqrt{2} G_F \sin \theta_c}
\frac{m_{K}^{2}}{(m_s + m_u) m_\mu}.
\label {eq:deltap}
\end{equation}

{}From this amplitude a partial decay width and transverse polarization
of the muon are calculated.  Since the effect of the $G_P$ coupling
is just to replace the coupling constant in the $M_{IB}$ term by a
complex one, the calculation of the transverse
polarization essentially reduces to the old calculation of T-odd
asymmetry in this process where the structure-dependent term were
assumed to be complex numbers\cite{Smoes}.
The partial decay width is given by,

\begin{eqnarray}
\frac{d^2 \Gamma}{dxdy} & = & \rho (x,y),
\end{eqnarray}
\begin{eqnarray}
\rho(x,y) & = & A_{IB}|1+\Delta_P |^2f_{IB}+A_{INT}(1+Re\Delta_P)
((V+A)f_{INT}^+ +(V-A)f_{INT}^-) \nonumber \\
& &+A_{SD}\frac{1}{2} ((V+A)^2 f_{SD}^+ + (V-A)^2
f_{SD}^-),
\label {eq:rho}
\end{eqnarray}
where $x$ and $y$ are normalized energies of the photon and muon, ie.
$x= \frac{2E_{\gamma}}{m_K}$, $y= \frac{2E_\mu}{m_K}$, and
$A_{SD}$ etc. are defined by,
\begin{eqnarray}
A_{SD} & = & \frac{m_K^5}{32 \pi^2} \alpha G_F^2 \sin^2 \theta_c,\\
A_{IB} & = & 2 r_\mu (\frac{\sqrt{2} f_K}{m_K})^2 A_{SD},\\
A_{INT}& = & 2 r_\mu (\frac{\sqrt{2} f_K}{m_K}) A_{SD},
\end{eqnarray}
and $r_\mu = (\frac{m_\mu}{m_K})^2$. Functions
$f_{IB} (x,y)$ etc. are defined in the appendix B.
Using a unit vector  $\vec{n}_T=\vec{p}_{\gamma}\times \vec{p}_{\mu}
/|\vec{p}_{\gamma}\times \vec{p}_{\mu}|$,
the muon transverse polarization is defined as
\begin{equation}
P_{\perp}=\frac{d \Gamma(\vec{n}_T) - d \Gamma(-\vec{n}_T) }
{d \Gamma(\vec{n}_T) + d \Gamma(-\vec{n}_T)},
\label {eq:def}
\end{equation}
where $\vec{p}_{\gamma}$ and $\vec{p}_{\mu}$ are the photon and muon
momenta
in the $K^+$ rest frame and $d\Gamma(\pm \vec{n}_T)$  is the partial
decay width with the muon polarization $\pm \vec{n}_T$.
$P_{\perp}$ is given by
\begin{equation}
P_{\perp} = \sigma (x,y) \cdot Im \Delta_P,
\end{equation}
where
\begin{eqnarray}
\sigma(x,y) & = & -A_P \cdot \frac{ 2\sqrt{(1-y+r_\mu)
((1-x)(x+y-1)-r_\mu)}}{\rho(x,y)}\nonumber\\
&  &\cdot \{ (V+A) f_p^+ + (V-A)f_p^-\}.
\label {eq:sigma}
\end{eqnarray}
$A_p$ is defined by,
\begin{equation}
A_p = \sqrt{r_\mu}\frac{\sqrt{2}f_K}{m_K}A_{SD},
\end{equation}
and $f_p^\pm$ are given in the appendix B.

Next, for completeness,  we present the partial width and the
transverse muon polarization in $K^+ \rightarrow \mu^+ \nu \pi^0$ decay
\cite{Kpimunu}.
Contrary to the $K^+ \rightarrow \mu^+ \nu \gamma$ process,
this process is sensitive to the scalar coupling $G_S$.
The form factors $f_\pm$ are defined as,
\begin{equation}
<\pi^0 | \bar{s} \gamma^\mu u | K^+> = f_+ (p+q)^\mu +
f_-(p-q)^\mu,
\end{equation}
where p, q are the $K^+$ and $\pi^0$ momenta. Using the fact
$|f_+| \gg |f_-|$, we get for the partial width
\begin{eqnarray}
\frac{d^2\Gamma}{dxdy} & = & \rho_{\pi}(x,y),\label{eq;phopi}\\
\rho_{\pi}(x,y) & = & \frac{m_K^5}{128\pi^3} G_F^2 \sin^2 \theta_c
f_+^2 \nonumber\\
& & \{(3+ r_{\mu} -r_\pi -x-2y)(x+2y-1-r_\mu
-r_\pi)\nonumber \\
& & -(1+r_\pi +x)(1+ r_\pi -r_\mu -x)\}.
\end{eqnarray}
where $x$ and $y$ are normalized energies for the pion and muon in
the $K^+$ rest frame, i.e. $x=\frac{2E_{\pi}}{m_K}$ and
$y= \frac{2E_\mu}{m_K}$.
The muon transverse polarization defined by the Eq. (\ref{eq:def})
with $\vec{n}_T=\vec{p}_{\pi}\times \vec{p}_{\mu}
/|\vec{p}_{\pi}\times \vec{p}_{\mu}|$ where $\vec{p}_{\pi}$
is the pion momentum is given by
\begin{eqnarray}
P_{\perp} &=&  \sigma_\pi (x,y)\cdot Im \Delta_S,\\
\sigma_\pi (x,y) &=& 2 \sqrt{r_\mu} \nonumber
\end{eqnarray}
\begin{equation}
\cdot \frac{\sqrt{
(x^2-4r_\pi)(y^2-4r_\mu)- 4(1+r_\mu +r_\pi
+\frac{xy}{2} -x-y)^2}}
{(3 +r_\mu -r_\pi -x-2y)(x+2y-1- r_\mu -r_\pi)
- (1+r_\pi+x)(1+ r_\pi- r_\mu -x)},\nonumber\\
\end{equation}
where
\begin{equation}
Im \Delta_S  =  \frac{(m_K^2 - m_\pi^2) Im G_S^*}{(m_s-m_u) m_\mu
\sqrt{2}G_F \sin \theta_c},
\label {eq:deltas}
\end{equation}
and $r_\pi = \frac{m_\pi^2}{m_K^2}$.
In the calculation of the partial width in Eq. (\ref{eq;phopi}),
we have assumed that the V-A
contribution is dominant and have kept  only the $G_F$ term.
Note that the polarization does not depend on
the form factor $f_+$ in the limit of $(f_+) \gg (f_-)$.

The Dalitz plots and transverse muon polarizations for
$K^+ \rightarrow \mu^+ \nu \gamma$ and $K^+ \rightarrow \mu^+ \nu
\pi^0$ are shown in figure 1, and 2.
In the calculations for $K^+ \rightarrow \mu^+ \nu \gamma$
the values of $V+A$ and $V-A$ have to be specified.
These values can be obtained in the analysis of radiative semileptonic
kaon decay. However, the present experimental knowledge is not
enough to extract finite numbers to these quantities, so that
we use the values obtained from calculation in the
one loop chiral perturbation theory which are $V+A=-0.137,
V-A=-0.052 $\cite{Bijnens}.
In the
$K^+ \rightarrow \mu^+ \nu \gamma$ process, the polarization
effect is large in the region $0.3 \le x \le 0.8$.
Although the differential decay width is large in the limit of
soft photon
($x \rightarrow 0$) the polarization vanishes in this limit.
This is because the transverse polarization is caused by
interference of $M_P$ and $M_{SD}$ term, whereas the large branching
in the limit of $x \rightarrow 0$ is caused by the
$|M_{IB} + M_P|^2 $ term.  Therefore, the sensitivity to the
polarization is determined by the intermediate $x$ region and
average polarization in this region is given by
$P_{\perp} = (0.1 \sim 0.2) \times Im\Delta_P$ depending on the
experimental
cut.  This is compared to the corresponding average polarization
for the $K^+ \rightarrow \pi^0 \mu \nu$ process where we get
$P_{\perp} \sim 0.3 \times Im \Delta_S$.
This shows that $K^+ \rightarrow \mu^+ \nu \gamma$ process
has a comparable sensitivity to new physics and the both
processes give valuable informations.
In general multi-Higgs models, $Im G_S$ and $Im G_P$ are not
related, therefore two process gives independent informations.
On the other hand, if we restrict ourself to the three Higgs
model they are expressed by the same parameters of the model,
so that we are able to obtain specific predictions.

\section{ Transverse Muon Polarization in Three Higgs Model}

In this section we consider prediction of muon transverse
polarization in $K^+ \rightarrow \mu^+ \nu \gamma$
and $K^+ \rightarrow \mu^+ \nu \pi^0$ decay in the context of
the three Higgs model.
We show that taking account of present phenomenological constraints
the predictions for the two processes are strongly correlated,
therefore it is important to search for T-odd polarization in
both processes.

Here we assume that three different Higgs doublets can couple
to up-type, down-type quarks and leptons, respectively.
Details on this model can be found in Refs.\cite{Kpimunu,
Grossman}.
There are two physical
charged Higgs and one Goldstone mode and the mixing matrix
among them can contain a new physical phase.
The original Yukawa coupling of this model is given by,

\begin{equation}
{\cal L} = \bar{q}_L y_d d_R H_d + \bar{q}_L y_u u_R \tilde{H_u}
+ \bar{\ell} y_e e_R H_{\ell} + h.c..
\end{equation}
We assume that the vacuum expectation values are in general
complex. If we define

\begin{eqnarray}
H_d & = & e^{i \theta_1}\left( \begin{array}{c} H_d^+\\
(\upsilon_1+\rho_1+i\chi_1)\end{array} \right),\\
H_u & = & e^{i \theta_2}\left( \begin{array}{c} H_u^+\\
(\upsilon_2+\rho_2+i\chi_2)\end{array} \right),\\
H_\ell & = & e^{i \theta_3}\left( \begin{array}{c} H_e^+\\
(\upsilon_3+\rho_3+i\chi_3)\end{array} \right),
\end{eqnarray}
where $H_{ui} \equiv \epsilon_{ij} \tilde{H}_{uj}^*$,
the above three charged Higgs fields $H_d^+$, $H_u^+$, $H_e^+$
are related to mass-diagonalized states by the following
3 $\times$ 3 matrix.

\begin{equation}
\left( \begin{array}{c} \frac{H_u^+}{\upsilon_1}\\
\frac{H_u^+}{\upsilon_2}\\ \frac{H_e^+}{\upsilon_3}\\
\end{array} \right)
= \frac{1}{\upsilon}
\left( \begin{array}{lrr}  1 & \alpha_1 & \alpha_2 \\
1 & -\beta_1 & -\beta_2 \\ 1 & \gamma_1 & \gamma_2
\end{array} \right) \left( \begin{array}{c} G^+ \\ H_1^+ \\
H_2^+ \end{array} \right),
\end{equation}
where $\upsilon = \sqrt{\upsilon_1^2 +\upsilon_2^2 + \upsilon_3^2}$,
$G^+$ is the Goldstone mode and $H_i$ (i = 1,2) are
physical charged Higgs mode.  The couplings between fermions and
the charged Higgses are determined as follows.

\begin{equation}
{\cal L }= (2\sqrt{2}G_F)^{\frac{1}{2}}\sum_{i=1}^{2} \{
\alpha_i \bar{u}_L K M_D d_R H_i^+ + \beta_i \bar{u_R} M_U K
d_L H_i^+ + \gamma_i \bar{\nu}_L M_E e_R H_i^+ \} + h.c.,
\end{equation}
where K is the ordinary flavor mixing matrix for the quark sector.
$M_D, M_U$ and $M_E$ are diagonal down-type quark, up-type quark
and lepton mass matrix respectively.
For the complex coupling constants $\alpha_i, \beta_i$ and $\gamma_i$,
the following relations exist from the requirement of unitarity
of the mixing matrix.

\begin{equation}
\frac{I_m (\alpha_2 \beta_2^*)}{I_m (\alpha_1 \beta_1^*)} =
\frac{I_m (\alpha_2 \gamma_2^*)}{I_m (\alpha_1 \gamma_1^*)}=
\frac{I_m (\beta_2 \gamma_2^*)}{I_m (\beta_1 \gamma_1^*)}=
-1,
\end{equation}
{}From this Lagrangian we can derive four-Fermi interaction
constants $G_S$ and $G_P$.

\begin{eqnarray}
G_P & = & \sum_{i=1}^{2} \sqrt{2}G_F \sin \theta_c m_\mu
\frac{\gamma_i}{m_i^2}(m_u \beta_i^* - m_s \alpha_i^*),\\
G_S & = & \sum_{i=1}^{2} \sqrt{2}G_F \sin \theta_c m_\mu
\frac{\gamma_i}{m_i^2}(m_s \alpha_i^* + m_u \beta_i^*).
\end{eqnarray}
In the formulas of the previous section, the above expressions
should be substituted to the coupling constants $G_S$ and $G_p$.

Since the transverse muon polarizations in $K^+ \rightarrow \mu^+
\nu \gamma$ and $K^+ \rightarrow \mu^+ \nu \pi^0$
are sensitive to the $Im G_P$ and
$Im G_S$ respectively,  $Im \Delta_P$ and $Im\Delta_S$ defined
from Eqs. (\ref {eq:deltap}) and (\ref {eq:deltas}) are given by,

\begin{eqnarray}
Im \Delta_P & = & \frac{m_K^2}{m_s + m_u}\sum_{i=1}^{2}
Im\{\frac{\gamma_i}{m_i^2}(m_u \beta_i^* - m_s \alpha_i^*)\}
\nonumber \\
&\simeq & - m_K^2 (\frac{1}{m_1^2}-\frac{1}{m_2^2})
(Im \gamma_1\alpha_1^* - \frac{m_u}{m_s} Im \gamma_1
\beta_1^*),
\label {eq:imdelp}
\end{eqnarray}
\begin{eqnarray}
Im \Delta_S & = & - \frac{m_K^2 - m_\pi^2}{m_s-m_u} \sum_{i=1}^{2}
\frac{1}{m_i^2}Im \{\gamma_i (m_s \alpha_i^* + m_u \beta_i^*)\}
\nonumber \\
& \simeq & - m_K^2 (\frac{1}{m_1^2} - \frac{1}{m_2^2})
(Im \gamma_1 \alpha_1^* + \frac{m_u}{m_s}Im \gamma_1 \beta_1^*),
\label {eq:imdels}
\end{eqnarray}
where we have neglected $m_u$ term in the denominator
in Eqs.(\ref {eq:imdelp}) and (\ref {eq:imdels})
and $m_\pi^2$ term compared
to $m_K^2$ term in Eq. (\ref {eq:imdels})
and used the unitarity relation to rewrite $Im\gamma_2\alpha_2^*$
and $Im \gamma_2 \beta_2^*$ in terms of $Im\gamma_1\alpha_1^*$ and
$Im \gamma_1 \beta_1^*$.
If we assume $m_1^2 \le m_2^2$ then the polarization effect is
maximal when $m_2 \rightarrow \infty$, on the other hand,
it vanishes when $m_1 = m_2$.  The two measurements of the muon
polarization can put constraints on the coupling constraints
$Im\gamma_1\alpha_1^*$ and $Im \gamma_1 \beta_1^*$
for a given set of charged Higgs masses.

In order to determine sensitivity,
we first discuss current bounds on these parameters from other
processes. From now on we are concentrating on the case $m_1\ll m_2$
and see what kinds of constraints are obtained for the
coupling constants of the lighter charged Higgs.
We denote the lighter charged Higgs mass as $m_H$.
Among $\alpha_i, \beta_i, \gamma_i$, the coupling constant
$\beta_i$ is most severely constrained since it is
related to the top Yukawa coupling constant.  We use the result of the
analysis in Ref. \cite{Grossman}.  For the $Im \gamma_1 \beta_1$,
the present bound is given by a product of the bounds of
$|\gamma_1|$ and $|\beta_1|$.
The bound of $|\beta_1|$ is given by $B^0 - \bar{B}^0$ mixing,
the parameter of CP violating amplitude of $K$ decay ($\epsilon$)
and the $Z \rightarrow b\bar{b}$ vertex.
For $m_t > 140 GeV$,
$|\beta_1| < 1.3 \sim 2.0 $ corresponding to the charged Higgs mass
45 GeV $\sim$ 200 GeV.
The bound on $|\gamma_1|$ is given by $e-\mu$ universality in $\tau$
decay and by the perturbative bound:
\begin{equation}
 |\gamma_1| < min ( 1.93 m_H~GeV^{-1}, 340).
\end{equation}
Then, $Im \gamma\beta^*$ is bounded by,

\begin{equation}
|Im \gamma^*_1 \beta_1| < 110 \sim 650,
\end{equation}
depending on $m_H= 45 GeV \sim 200 $GeV.
The strongest bound on $Im \gamma \alpha^*$ is obtained from
$B \rightarrow  X\tau \nu_\tau$ decay in the range of $m_H < 400$
GeV.

\begin{equation}
|Im \gamma^*_1 \alpha_1 | < 0.23 (\frac{m_H}{GeV})^2,
\end{equation}
which varies from 465 to 9200 for the range of $m_H$ from 45 GeV
 to 200 GeV.
Note that in the two expressions, $Im \Delta_P$ and $Im\Delta_S$,
the second term $\frac{m_u}{m_s}Im\gamma_1\beta_1^*$ is more strongly
constrained than the first term $Im \gamma_1 \alpha_1^*$.

In figure 3, we show constraints on $Im \gamma_1 \beta_1^*$ and
$Im \gamma_1 \alpha_1^*$ space expected from future muon polarization
measurements for different values of $m_H$. We have assumed $P_{\perp}
=0.2 \times Im\Delta_P$ and $P_{\perp}
=0.3 \times Im\Delta_S$ for the $K^+ \rightarrow \mu^+
\nu \gamma$ and $K^+ \rightarrow \mu^+ \nu \pi^0$ processes
respectively.
In the analysis we have used $V+A=-0.137, V-A=-0.052 $ as before
and $\frac{m_u}{m_s}=\frac{1}{40}$.
The bounds from these two processes are presented.  Also the present
experimental constraints from other processes are shown.

{}From the figure 3, we can see that the both processes are quite
useful to put constraints on the value of $Im \gamma_1 \alpha_1^*$.
Also a strong correlation between the
prediction of the two polarizations can be seen.
This is because $Im \gamma_1 \beta_1^*$ is already strongly constrained
from other precesses.
Therefore, if the coming experiment gives non-null result for the
polarization
measurements, the $Im \gamma_1 \alpha_1^*$ term will be dominant and
the prediction of two polarizations are strongly correlated.
This is important for the experiment because if the T-odd
polarization is observed in one process then the polarization
in the other process is also expected to be within reach.
Notice that this strong correlation is a unique feature of this three
Higgs model where the parameter $Im \gamma_1^* \beta_1$ is strongly
bounded because the coupling constant $\beta_i$ is related to the
processes involving top Yukawa coupling. If we allow more Higgses,
the predictions for two polarizations are not necessarily correlated.

\section{Discussions}

We have calculated the partial decay width and muon transverse
polarization in the processes $K^+ \rightarrow \mu^+ \nu \pi^0$
and $K^+ \rightarrow \mu^+ \nu \gamma$ in the model with complex
scalar and pseudo scalar couplings.
For the calculation for the $K^+ \rightarrow \mu^+ \nu \gamma$
process we do not need to introduce any new form factor,
therefore, no new theoretical ambiguity exists to extract
short distance effects of new physics. Improvements on the
polarization measurements expected at the new experiment will
give remarkable impacts on the search for a new source of CP
violation in the Higgs sector. Especially in the three
Higgs model we have shown that the predictions of two polarization
are strongly correlated, therefore it is important to search for
T-violation in both processes.

In the actual experiment, the final state interaction due to
the electromagnetic interaction induces a T-odd effect which mimics
the T-violation.
For the $K^+ \rightarrow \mu^+ \nu \pi^0$ process, this effect is
evaluated to be $10^{-6}$ and therefore negligible.
On the other hand for the $K^+ \rightarrow \mu^+ \nu \gamma$
process the final state interaction can induce the muon
transverse polarization at the level of $10^{-3}$.
At first sight this looks a problem for measuring the T-violating
 effects in this process. This is, however, not the case
since the effect is induced by the electromagnetic interaction
and can be estimated without much ambiguity.  Moreover, in a very
good approximation, the total transverse polarization is expected
to be a simple sum of a term due to the final state interaction
and that from the pseudo scalar coupling since each term comes
from the interference with the standard model tree amplitude.
Therefore, the subtraction procedure is straightforward.
Detailed evaluation of the final state interaction for
this process is called for.

The authors would like to thank Y. Kuno for valuable discussions.
The work of Y.O. is in part supported by the Grant-in-aid for
Scientific Research from the Ministry of Education
and Culture of Japan. The work of T. T. L. is supported by
Nishina Memorial Foundation.

\newpage
\appendix
\section{Appendix A}
In this appendix we derive the formula (\ref {eq:M_P})
for the amplitude of $K^+ \rightarrow \mu^+ \nu \gamma$
from the pseudo scalar and scalar couplings.
There are two types of diagrams for this process from the
scalar and pseudo scalar couplings defined in Eq. (\ref {eq:lag}).
The first diagram is the one in which the photon is originated
from the external muon line. We get the following amplitude for
this diagram,
\begin{equation}
M_{1}  =  e  \frac{G_P}{2}
< 0  | \bar{s} \gamma^5 u | K^+(p)>
\epsilon^*_{\mu}(q)\bar{u}(k)(1+\gamma_5)(
\frac{q\cdot \gamma \gamma^\mu + 2 \ell^\mu}{2 \ell \cdot q})
\upsilon (\ell).
\label {eq:M_1}
\end{equation}
In the above expression the contribution from $G_S$ has dropped
because of parity conservation in the matrix element.

The second diagram corresponds to the situation in which the photon
comes from the hadronic system. These contribution can be written as,
\begin{equation}
M_{2}  =  -ie \frac{G_P}{2}
\epsilon^*_{\mu}(q)\bar{u}(k)(1+\gamma_5) \upsilon (\ell)
(I^{\mu}_S+I^{\mu}_P),
\end{equation}
where
\begin{eqnarray}
I^{\mu}_S &=&
\int d^4 x e^{iqx}<0 | T(\bar{s} u(0) J_{em}^{\mu}(x))| K^+(p)>,\\
I^{\mu}_P &=&
\int d^4 x e^{iqx}<0 | T(\bar{s}\gamma^5 u(0)
J_{em}^{\mu}(x))| K^+(p)>.
\end{eqnarray}
Here, $I^{\mu}_S$ and $I^{\mu}_P$ are functions of two momenta $p^{\mu},
q^{\mu}$. Since we cannot construct axial vector quantity from
two independent momenta  we can set  $I^{\mu}_S=0$. On the
other hand, $I^{\mu}_P$ can be expanded as
\begin{equation}
I^{\mu}_P=I_1\cdot p^{\mu}+I_2 \cdot q^{\mu}.
\end{equation}
The $I_2$ part does not contribute to the on-shell photon amplitude.
The $I_1$ is determined using the Ward-Takahashi identity. In fact
we can show
\begin{equation}
q_{\mu} I^{\mu}_P=-i< 0  | \bar{s} \gamma^5 u | K^+(p)>.
\end{equation}
Then,
\begin{equation}
I^{\mu}_P=-i\frac{p^{\mu}}{p\cdot q}< 0  | \bar{s}
\gamma^5 u | K^+(p)>.
\end{equation}
Therefore, the second amplitude is written as follows;
\begin{equation}
M_{2}  = - e  \frac{G_P}{2}
< 0  | \bar{s} \gamma^5 u | K^+(p)>(\frac{p^{\mu}}{p\cdot q})
\epsilon^*_{\mu}(q)\bar{u}(k)(1+\gamma_5)\upsilon (\ell).
\label {eq:M_2}
\end{equation}
Combining Eqs. (\ref{eq:M_1}) and (\ref{eq:M_2}), and expressing
the scalar matrix element by the decay constant as
\begin{equation}
< 0  | \bar{s} \gamma^5 u | K^+(p)>=
i\frac{\sqrt{2}f_{K}m_{K}^{2}}{m_s + m_u},
\end{equation}
we obtain Eq. (\ref{eq:M_P}).
\section{Appendix B}
In this appendix the functions for evaluation of the partial
width and transverse muon polarization for the
$K^+ \rightarrow \mu^+ \nu \gamma$ process are listed.

\begin{eqnarray}
f_{IB}(x,y) & = & \frac{1-y+r_\mu}
{x^2(x+y-1-r_\mu)}
(x^2+2(1-x)(1-r_\mu) -\frac{2xr_\mu(1-r_\mu)}
{x+y-1-r_\mu}),\\
f_{SD}^+(x,y) & = & (x+y-1-r_\mu)((x+y-1)(1-x) -r_\mu),\\
f_{SD}^-(x,y) & = & (1-y+r_\mu)((1-x)(1-y) + r_\mu),\\
f_{INT}^+(x,y) & = & (\frac{1-y+r_\mu}{x(x+y-1-r_\mu)})
((1-x)(1-x-y) + r_\mu),\\
f_{INT}^-(x,y) & = & (\frac{1-y+ r_\mu}{x(x+y-1-r_\mu)})
(x^2 -(1-x)(1-x-y) -r_\mu),\\
f_p^+(x,y) & = & \frac{(1-x)(x+y-1) -r_\mu}{x(x+y-1-r_\mu)},\\
f_p^-(x,y) & = & \frac{1+r_\mu -y}{x(x+y-1-r_\mu)}.\\
\end{eqnarray}
In the above, $x$ and $y$ are defined as $x=\frac{2E_{\gamma}}{m_K}$
and $y= \frac{2E_\mu}{m_K}$ using the photon and muon energies
in the $K^+$ rest frame.
\newpage
\noindent{\large {\bf Figure Captions}}\\
\\
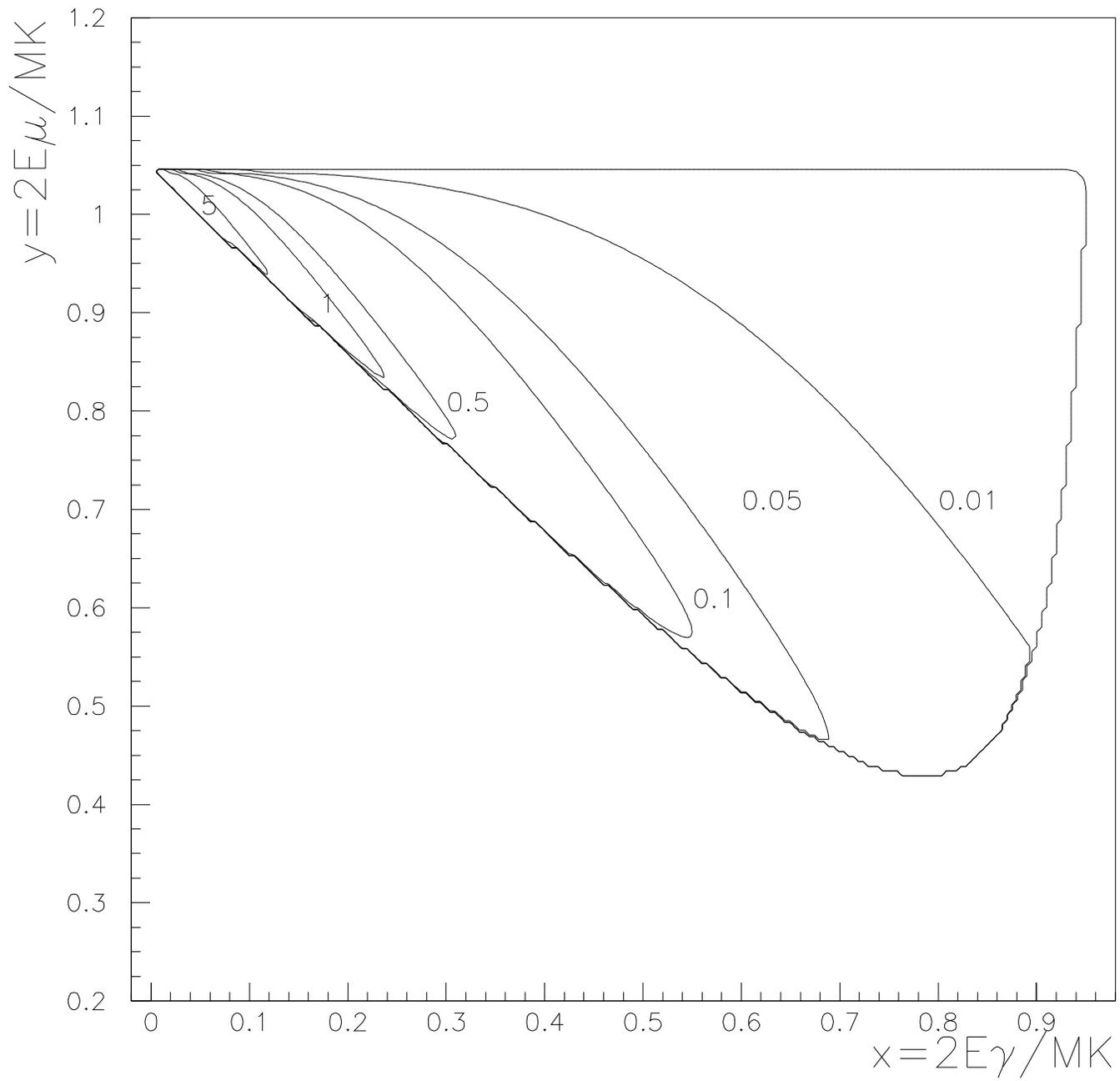
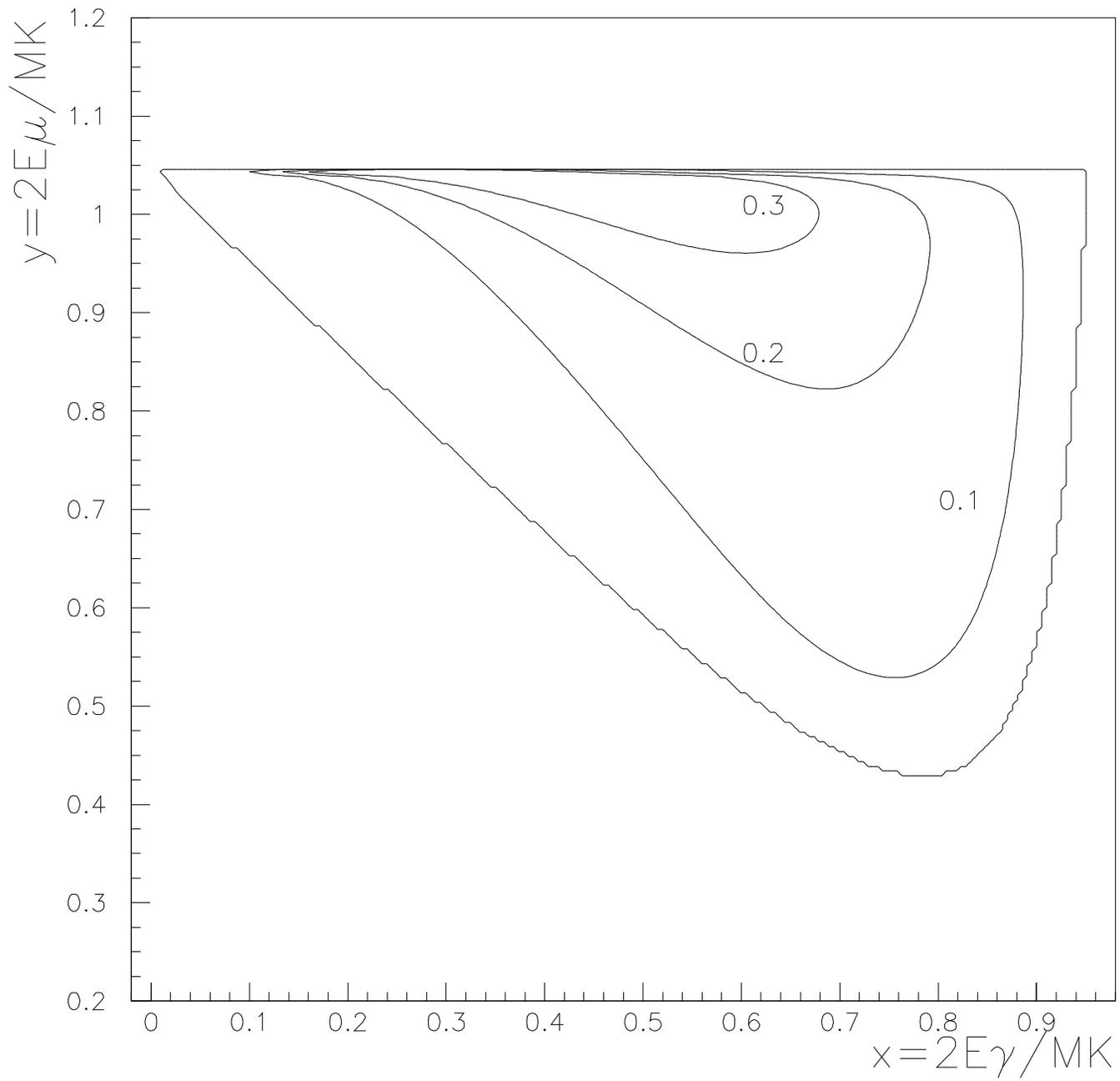
{\bf Figure 1} ~~Partial decay width (a) and transverse muon
polarization (b) for
the $K^+ \rightarrow \mu^+ \nu \gamma$ process
as a function of $x=\frac{2E_{\gamma}}{m_K}$ and
$y= \frac{2E_\mu}{m_K}$. (a) represents the partial decay width
$\rho(x,y)$ normalized by a constant $A_{SD}$ and  (b)
represents the function $\sigma(x,y)$.
\\

\noindent
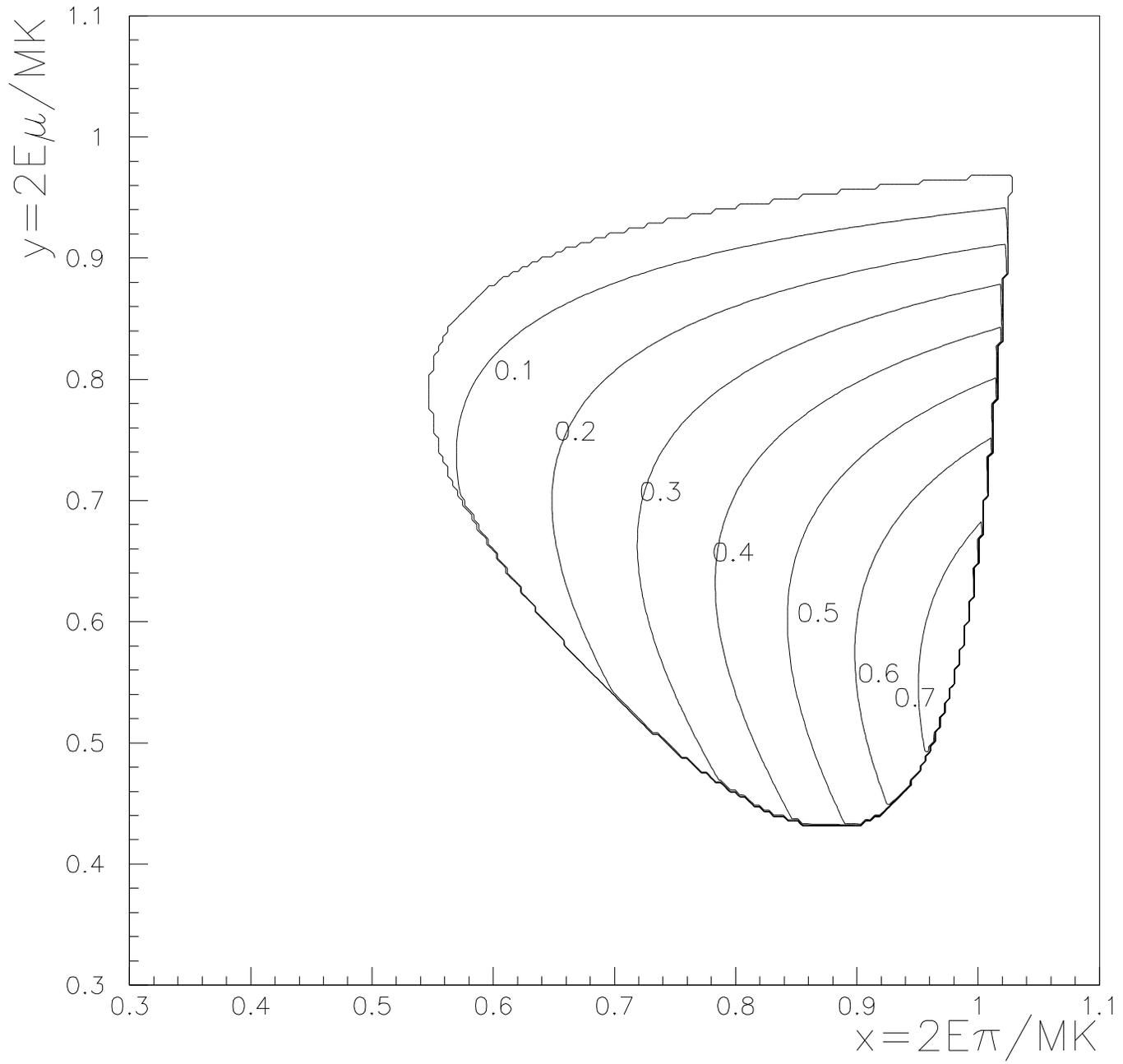
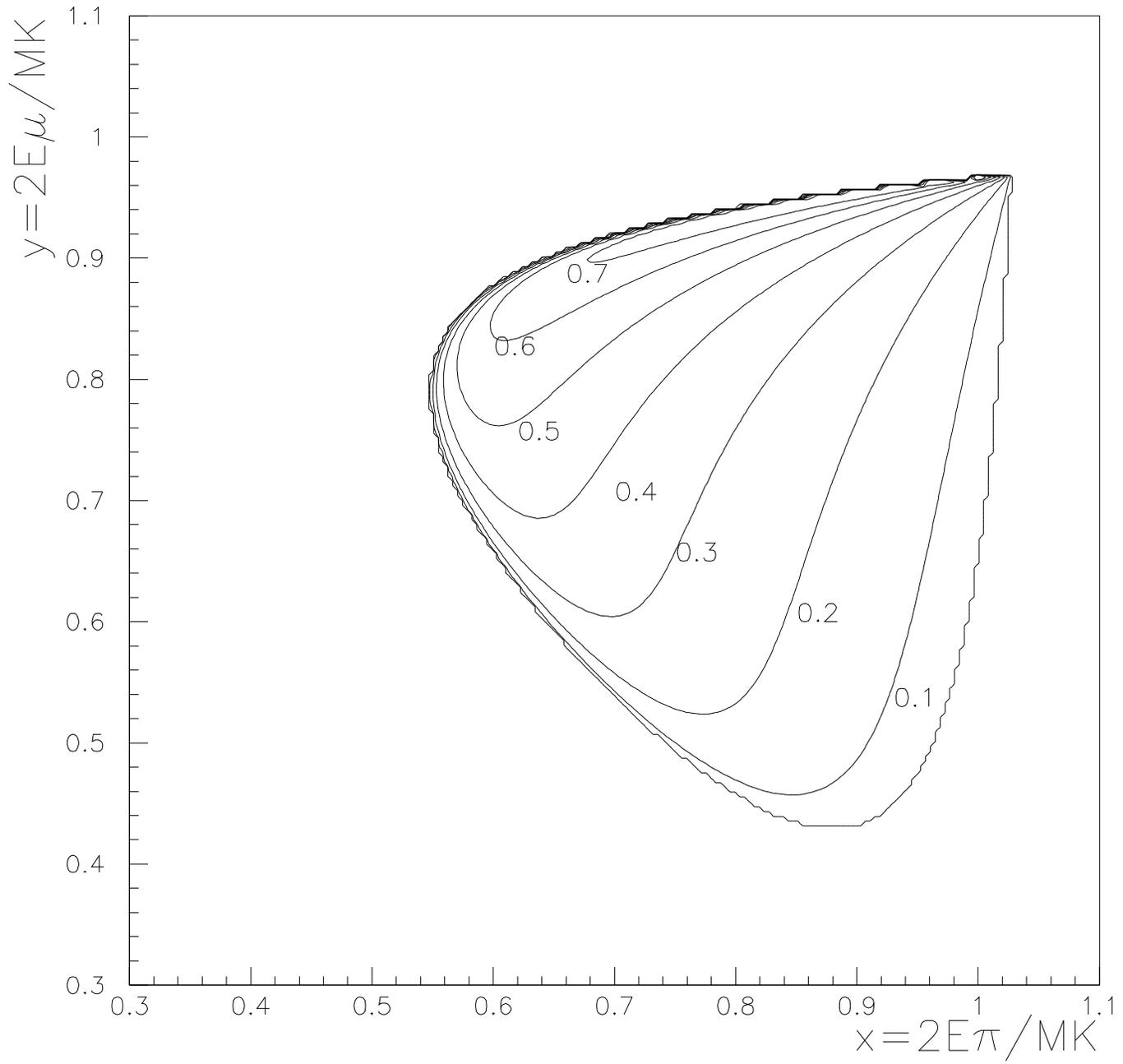
{\bf Figure 2} ~~Partial decay width (a) and transverse muon
polarization (b) for
the $K^+ \rightarrow \mu^+ \nu \pi^0$ process as a function of
$x=\frac{2E_{\pi}}{m_K}$ and $y= \frac{2E_\mu}{m_K}$. (a) represents
the partial decay width
$\rho_{\pi}(x,y)$ normalized by  $\frac{m_K^5}{128\pi^3}
G_F^2 \sin^2 \theta_c f_+^2 $ and  (b) represents the function
$\sigma_{\pi}(x,y)$.\\

\noindent
{\bf Figure 3} ~~Constraints on the parameters of the three
Higgs model obtained from transverse muon polarization measurements
for the charged Higgs mass
45 GeV (a) and 200 GeV (b). The solid (dotted) lines
correspond to the $K^+ \rightarrow \mu^+ \nu \gamma$
($K^+ \rightarrow \mu^+ \nu \pi^0$) process. From left to right
the lines represent  $P_{\perp}=5\times 10^{-3},
1\times 10^{-3},5\times 10^{-4},-5\times 10^{-4},-1\times 10^{-3},
-5\times 10^{-3}$ for both cases.
The shaded parameter regions are
already excluded by other phenomenological constraints.

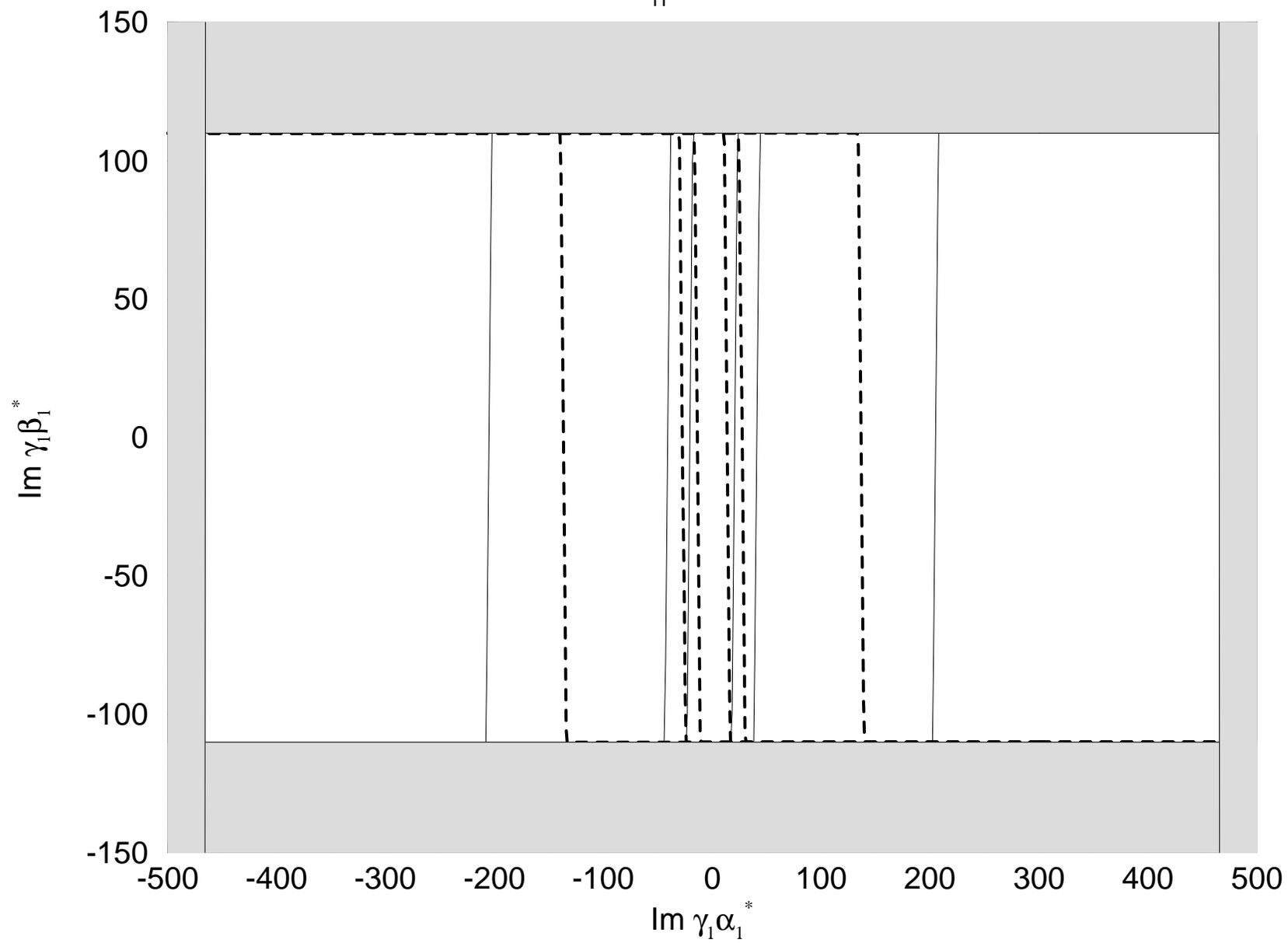
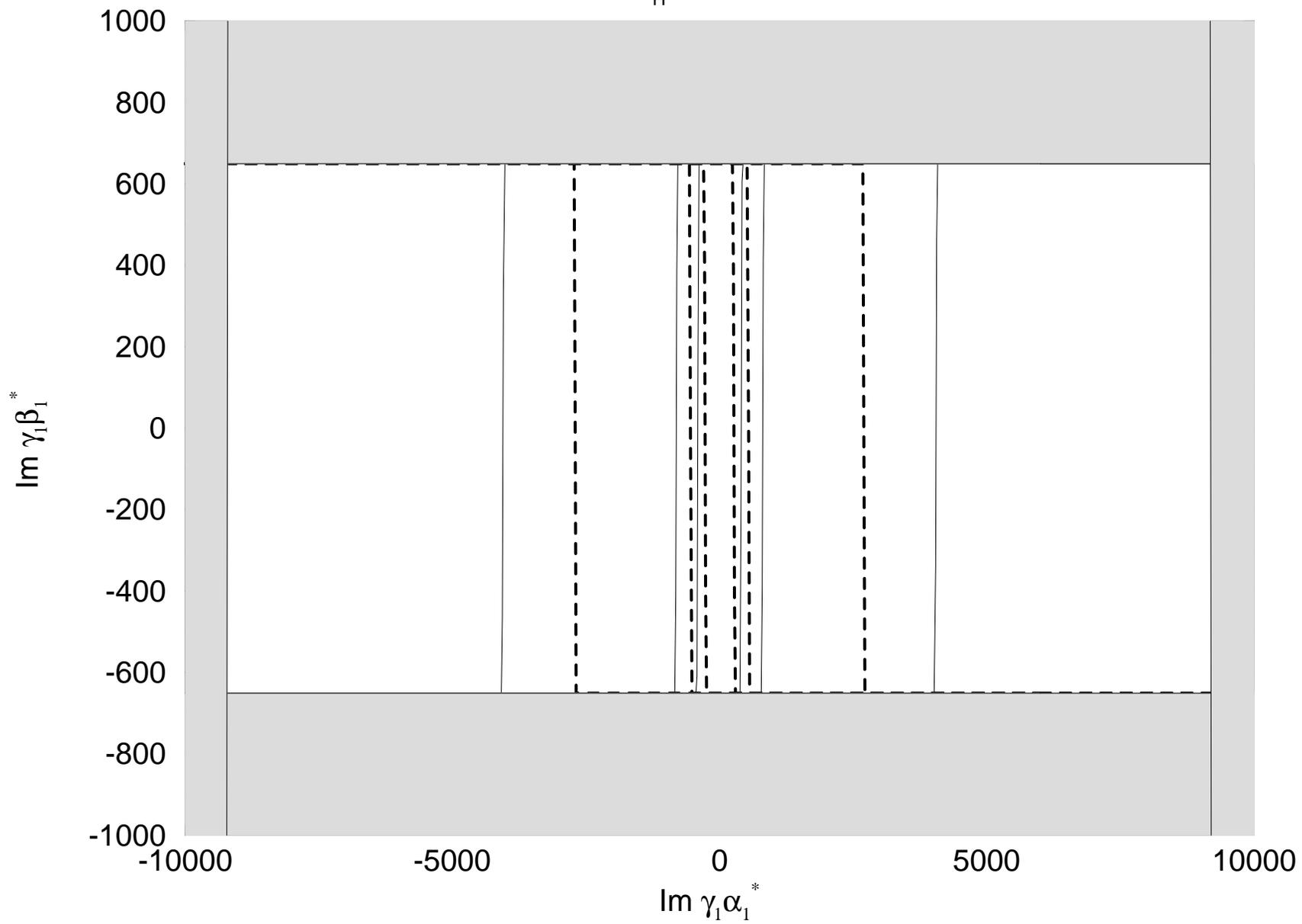
\end{document}